# Full transmission within a wide energy range and super-criticality in relativistic barrier scattering

A. D. Alhaidari[1], H. Bahlouli[1,2], Y. Benabderahmane[3] and A. Jellal[1,3]

[1]*Saudi Center for Theoretical Physics, Dhahran, Saudi Arabia*
[2]*Physics Department, King Fahd University of Petroleum & Minerals, Dhahran 31261, Saudi Arabia*
[3]*Theoretical Physics Group, Faculty of Sciences, Chouaïb Doukkali University, PO Box 20, 24000 El Jadida, Morocco*

**Abstract**: For potential barriers with scalar and vector coupling, we show that a Dirac particle could experience nearly full transmission within a wide sub-barrier energy band. Moreover, for certain potential configurations, including pseudo-spin symmetry where the scalar potential is the negative of the vector, full transmission occurs for arbitrarily small momentum.

## I. Introduction

The possibility of a quantum particle penetrating an energetic barrier, called tunneling, has been recognized as one of the hallmarks of quantum theory and represents certainly one of the most spectacular consequences of quantum mechanics [1]. Tunneling is a purely quantum phenomenon that happens in the classically forbidden region; its experimental observation constituted a very important support to the quantum theory. In fact, tunneling phenomena played an important role in nonrelativistic quantum mechanics due to its important application in electronic devices. It was Leo Esaki who discovered a characteristic called negative differential resistance (NDR) whereby, for PN junction diodes, the current voltage characteristics has a sharp peak at a certain voltage associated with resonant tunneling. This constituted the first important confirmation that this phenomenon is due to the quantum mechanical tunneling effect of electrons [2]. These observations led to various applications in atomic and molecular physics as well as in mesoscopic science (for instance in the context of spintronics devices [3], to mention just one recent example). Transmission resonances occur when the transmission coefficient reaches unity, thus resonances give rise to a maximum or peak in a certain measure of the scattering such as the transmission coefficient in one dimension or the cross-section for a particular partial wave in three dimensions. However, Wigner [4] pointed out a long time ago that in order to have a true physical resonance the scattered particle must have a phase shift associated with the peak that increases through an odd multiple of $\pi/2$ as the particle energy increases through the resonance. Usually when an electron is incident on a complex potential barrier/well structure and has an energy that coincides with one of the bound states of the potential then resonant tunneling occurs and the electron can be transmitted with a transmission coefficient of the order of unity.

In addition to these usual resonances at finite energy, zero energy resonance can also occur when the potential supports a bound state of zero energy. This bound state can take the form $E = \frac{-1}{2m}p^2$ with the momentum $p$ vanishingly small, $p \to 0$, the corresponding wavefunction thus becomes a continuum wavefunction. This zero-energy resonance is called a half-bound state since the corresponding wavefunction is finite but does not decay fast enough at infinity to ensure square integrability [5].

The study of tunneling of relativistic particles through one-dimensional potentials has been performed for some limited and simple potential configurations, such as potential barriers and the analysis of resonant tunneling through multi-barrier systems [6]. However, recent advances in graphene technology [7] made it possible to study the physics of relativistic electrons in graphene based solid state devices, whose behavior differs drastically from that of similar devices fabricated with usual semiconductors. Consequently, new unexpected phenomena have been observed while other phenomena that were well-understood in common semi-conductors, such as the quantum Hall effect and weak-localization, exhibited surprising behavior in graphene. Thus, graphene devices enabled the study of relativistic dynamics in controllable nano-electronic circuits and allowed for the observation of some subtle effects, previously accessible only to high energy physics, such as Klein tunneling. Recently, electron transport through electrostatic barriers in single and bi-layer graphene has been studied using the Dirac equation [8]. The study of transmission resonances in relativistic wave equations in external potentials has been discussed extensively in the literature [9-10]. In this case, for given values of the energy and shape of the barrier, the probability of transmission reaches unity even if the potential strength is larger than the energy of the particle, a phenomenon that is not present in the nonrelativistic case. The relation between low momentum resonances and super-criticality has been established by Dombey *et al.* [11] and Kennedy [12]. Some results on the scattering of Dirac particles by a one-dimensional potential exhibiting resonant behavior have also been reported [10-13].

In this manuscript, we will be interested mainly in, the so-called, supercritical transmission of a relativistic particle through potential barriers. That is, a Dirac particle with arbitrarily small momentum experiences full transmission through a square potential barrier with scalar and vector coupling. Another objective of our study is to find the right configuration in such potential barriers so that a Dirac particle could experience nearly full transmission within a wide sub-barrier energy band. Now, the energy spectrum of a relativistic particle is very peculiar. In fact, for a free Dirac particle, there exists a gap $|\varepsilon| \leq mc^2$ which separates the positive and negative energy continuum states [14]. The positive energy states correspond to particle states and the absence of negative energy states (hole states) describe anti-particles. The introduction of a potential $V(x)$ creates a distortion in the gap and bound states may now occur between $\varepsilon = -mc^2$ and $\varepsilon = +mc^2$. As a result of this energy dispersion, the Dirac equation has half-bound states at both $\varepsilon = -mc^2$ and $\varepsilon = +mc^2$ in contrast to the Schrödinger equation where there exist only one half-bound state at $E = 0$. Moreover, it will be more suitable to talk about zero momentum resonances in the relativistic case rather than zero energy resonances. The objective of our present work is to design the right potential configuration in the 1+1 Dirac equation which will enable us to achieve one of two goals:
(1) Meet the supercritical condition where full transmission occurs for arbitrarily small momentum.
(2) Allow for nearly full transmission within a sub-barrier energy band, which is very wide, in the Klein energy zone.

Due to the fact that the potential interaction in the Dirac equation has a matrix form rather than a scalar form, this will result in more degrees of freedom while selecting the coupling mode which will give rise to the two above listed phenomena. In this work, we show that the Dirac equation in 1+1 space-time dimension with scalar and vector coupling, including pseudo-spin symmetric configuration, will give rise to supercriticality. This latter type of configuration attracted a lot of attention in the literature due to the resulting simplification in the solution of the relativistic problem. The wave equation, in this case, could always be reduced to a Schrödinger-type second order differential equation. This puts at one's disposal a variety of well-established analytic tools and techniques to be employed in the analysis and solution of the problem [15]. However, we made a thorough analysis of this potential configuration and gave a proper physical interpretation of this class of problems [16]. Moreover, for a suitable combination of a vector potential barrier and a scalar potential well (a generalization of the pseudo-spin symmetric case) the second objective of our study could be achieved where full transmission occurs within a very wide sub-barrier energy band in the Klein energy zone.

The manuscript is organized as follows. In section II, we introduce the Dirac equation in 1+1 dimensions and identify the different parameters that affect our problem. In section III, we solve explicitly the Dirac equation for a combination of vector and scalar potential barriers and obtain the eigenspinor solutions. Finally, in section IV, we give explicit results of our findings and close by a discussion and conclusion.

**II. Dirac equation in 1+1 space-time**

To describe the interaction in the Dirac equation it is not enough just to give a potential function, one has also to identify the coupling mode. This is easily understood by noting that the Dirac Hamiltonian in 1+1 dimension is a 2×2 matrix. Thus, while maintaining hermiticity, one can include a real potential function in this Hamiltonian in one of four alternative ways: $V(x) \times I$, $W(x) \times \sigma_1$, $U(x) \times \sigma_2$, or $S(x) \times \sigma_3$, where $I$ is the 2×2 unit matrix and the $\sigma$'s are the usual three Pauli matrices. Consequently, the potential interaction in the 1+1 Dirac equation can be in one of three modes: vector, scalar, and pseudo-scalar. That is, $(V,U)$, $W$, and $S$ are the vector, pseudo-scalar and scalar couplings, respectively. These potential components are named after their transformation properties under Lorentz transformation. Using $U(1)$ gauge invariance, we can eliminate the space component of the vector potential, $U \times \sigma_2$. Thus, the time independent Dirac equation for a point particle of rest mass $m$ in 1+1 space-time dimension with the most general potential coupling reads as follows

$$\begin{pmatrix} +m + S(x) + V(x) - \varepsilon & -\frac{d}{dx} + W(x) - iU(x) \\ +\frac{d}{dx} + W(x) + iU(x) & -m - S(x) + V(x) - \varepsilon \end{pmatrix} \begin{pmatrix} \psi^+(x) \\ \psi^-(x) \end{pmatrix} = 0, \qquad (1)$$

where we have used the conventional relativistic units $\hbar = c = 1$. Now, with $U = W = 0$, the equation becomes

$$\begin{pmatrix} +m + V_+(x) - \varepsilon & -\frac{d}{dx} \\ +\frac{d}{dx} & -m + V_-(x) - \varepsilon \end{pmatrix} \begin{pmatrix} \psi^+(x) \\ \psi^-(x) \end{pmatrix} = 0, \qquad (2)$$

where $V_\pm = V \pm S$ and we obtain the following coupled differential relations between the two components of the wavefunction

$$\psi^+ = \frac{1}{m - \varepsilon + V_+} \frac{d\psi^-}{dx}, \qquad (3a)$$

$$\psi^- = \frac{1}{m+\varepsilon-V_-}\frac{d\psi^+}{dx}. \tag{3b}$$

Substituting the first into the second, we obtain

$$\left[\frac{d^2}{dx^2}+(\varepsilon-m-V_+)(\varepsilon+m-V_-)+\frac{dV_+/dx}{\varepsilon-m-V_+}\frac{d}{dx}\right]\psi^- = 0. \tag{4a}$$

On the other hand, if we substitute the second into the first we get

$$\left[\frac{d^2}{dx^2}+(\varepsilon-m-V_+)(\varepsilon+m-V_-)+\frac{dV_-/dx}{\varepsilon+m-V_-}\frac{d}{dx}\right]\psi^+ = 0. \tag{4b}$$

Solving (4a) and substituting that solution into (3a) gives the two-component wavefunction associated with one solution space. However, solving (4b) and substituting that solution into (3b) gives another two-component wavefunction belonging to a second solution space. Below we show that there is no ambiguity in the solution of the problem and that both solution spaces are required. In fact, the total solution space consists of the union of these two subspaces whose individual validity depends on the value of the system's energy.

**III. Solution spaces for square barrier/well potentials**

Let us assume that the potential functions $V(x)$ and $S(x)$ are piecewise constant. Thus, within regions where the potentials are constant we obtain

$$\left[\frac{d^2}{dx^2}+(\varepsilon-m-V_+)(\varepsilon+m-V_-)\right]\psi^\pm = \left[\frac{d^2}{dx^2}+(\varepsilon-V)^2-(m+S)^2\right]\psi^\pm = 0, \tag{5}$$

and the two boundaries of the continuous energy spectrum are at $\varepsilon = V \pm (m+S)$. Moreover, positive/negative energy solutions correspond to $\varepsilon$ greater/less than $V$. Additionally, we end up with two distinct cases depending on whether $S$ is greater than or less than $-m$.

<u>Case 1</u>: $S > -m$ :

Positive Energy ($\varepsilon > V$):  $\psi = \begin{pmatrix} \psi^+ \\ \dfrac{1}{m+\varepsilon-V_-}\dfrac{d\psi^+}{dx} \end{pmatrix}.$ \hfill (6a)

Negative Energy ($\varepsilon < V$):  $\psi = \begin{pmatrix} \dfrac{1}{m-\varepsilon+V_+}\dfrac{d\psi^-}{dx} \\ \psi^- \end{pmatrix}.$ \hfill (6b)

<u>Case 2</u>: $S < -m$ :

Positive Energy ($\varepsilon > V$):  $\psi = \begin{pmatrix} \dfrac{1}{m-\varepsilon+V_+}\dfrac{d\psi^-}{dx} \\ \psi^- \end{pmatrix}.$ \hfill (7a)

Negative Energy ($\varepsilon < V$):  $\psi = \begin{pmatrix} \psi^+ \\ \dfrac{1}{m+\varepsilon-V_-}\dfrac{d\psi^+}{dx} \end{pmatrix}.$ \hfill (7b)

The solution of (5) is

$$\psi^{\pm}(x) = A_{\pm} e^{+px} + B_{\pm} e^{-px},$$

where $p^2 = (m+S)^2 - (\varepsilon - V)^2$ and we obtain the following solutions for $S > -m$

$$\varepsilon > V: \quad \psi = \begin{pmatrix} 1 \\ \alpha \end{pmatrix} A_+ e^{+px} + \begin{pmatrix} 1 \\ -\alpha \end{pmatrix} B_+ e^{-px}. \tag{6a}'$$

$$\varepsilon < V: \quad \psi = \begin{pmatrix} \beta \\ 1 \end{pmatrix} A_- e^{+px} + \begin{pmatrix} -\beta \\ 1 \end{pmatrix} B_- e^{-px}, \tag{6b}'$$

where $\alpha = p/(m + \varepsilon - V_-)$ and $\beta = p/(m - \varepsilon + V_+) = \alpha^{-1}$. Moreover, oscillatory solutions are obtained for $\varepsilon > V_+ + m$ and for $\varepsilon < V_- - m$. On the other hand, for $S < -m$, we obtain

$$\varepsilon > V: \quad \psi = \begin{pmatrix} \beta \\ 1 \end{pmatrix} A_- e^{+px} + \begin{pmatrix} -\beta \\ 1 \end{pmatrix} B_- e^{-px}. \tag{7a}'$$

$$\varepsilon < V: \quad \psi = \begin{pmatrix} 1 \\ \alpha \end{pmatrix} A_+ e^{+px} + \begin{pmatrix} 1 \\ -\alpha \end{pmatrix} B_+ e^{-px}, \tag{7b}'$$

and oscillatory solutions are obtained for $\varepsilon > V_- - m$ and for $\varepsilon < V_+ + m$.

In the following, we assume scattering off a square potential barrier/well. The height/depth of the barrier/well for vector and scalar coupling is $V$ and $S$, respectively. From now on, we refer to both barrier and well simply as "barrier". The edges for both barriers are located at $x = 0$ and $x = a$. Now, outside the barrier, the solution of the free Dirac equation is

$$x < 0: \quad \psi(x) = \begin{pmatrix} 1 \\ i\gamma \end{pmatrix} e^{ikx} + \begin{pmatrix} 1 \\ -i\gamma \end{pmatrix} R e^{-ikx}, \tag{8}$$

$$x > a: \quad \psi(x) = \begin{pmatrix} 1 \\ i\gamma \end{pmatrix} T e^{ikx}, \tag{9}$$

where $k = \sqrt{\varepsilon^2 - m^2}$, $\gamma = \sqrt{(\varepsilon - m)/(\varepsilon + m)}$ and $\varepsilon \geq m$. We have normalized the flux of the incident beam from left to unit amplitude and $R$ is the reflection amplitude whereas $T$ is the transmission amplitude. By equating the spinor wavefunction at the boundaries $x = 0$ and $x = a$ we obtain four equations with four unknowns ($R, T, A_\pm, B_\pm$).

Now, for $S > -m$ the positive/negative energy solutions inside the barrier are given by (6a)'/(6b)'. The necessary, but not sufficient, condition for sub-barrier full transmission is that the positive energy solutions outside the barrier overlap with the negative energy solution inside the barrier. That is, if we take $V_- - m \geq m$; equivalently, $V - 2m \geq S > -m$. Solving for the reflection and transmission amplitudes, we obtain:

$$\varepsilon > V: \quad R(\varepsilon) = \frac{i\gamma - \alpha}{i\gamma + \alpha}(1 - e^{2pa})\left[1 - e^{2pa}\left(\frac{i\gamma - \alpha}{i\gamma + \alpha}\right)^2\right]^{-1}. \tag{10a}$$

$$T(\varepsilon) = e^{a(p-ik)}\left[1 - \left(\frac{i\gamma - \alpha}{i\gamma + \alpha}\right) R(\varepsilon)\right]$$

$$= e^{a(p-ik)} \frac{4i\alpha\gamma}{(i\gamma + \alpha)^2}\left[1 - e^{2pa}\left(\frac{i\gamma - \alpha}{i\gamma + \alpha}\right)^2\right]^{-1} \tag{11a}$$

$$\varepsilon < V: \quad R(\varepsilon) = \frac{i\beta\gamma - 1}{i\beta\gamma + 1}(1 - e^{2pa})\left[1 - e^{2pa}\left(\frac{i\beta\gamma - 1}{i\beta\gamma + 1}\right)^2\right]^{-1}. \tag{10b}$$

$$T(\varepsilon) = e^{a(p-ik)} \left[ 1 - \left( \tfrac{i\beta\gamma-1}{i\beta\gamma+1} \right) R(\varepsilon) \right]$$
$$= e^{a(p-ik)} \frac{4i\beta\gamma}{(i\beta\gamma+1)^2} \left[ 1 - e^{2pa} \left( \tfrac{i\beta\gamma-1}{i\beta\gamma+1} \right)^2 \right]^{-1} \quad (11b)$$

Now, since $\beta = \alpha^{-1}$ the above expressions for the scattering amplitudes are the same for $\varepsilon > V$ and $\varepsilon < V$. Moreover, these expressions could be rewritten for all energies, $\varepsilon \geq m$, in the following compact notation:

$$R(\varepsilon) = \mu(\varepsilon)(1 - e^{2pa}) \left[ 1 - e^{2pa}\mu(\varepsilon)^2 \right]^{-1}. \quad (12)$$

$$T(\varepsilon) = e^{a(p-ik)} [1 - \mu(\varepsilon) R(\varepsilon)] = e^{a(p-ik)} [1 - \mu(\varepsilon)^2] \left[ 1 - e^{2pa}\mu(\varepsilon)^2 \right]^{-1}. \quad (13)$$

where the energy function ratio $\mu(\varepsilon) = (i\gamma - \alpha)/(i\gamma + \alpha)$. One can show that $|\mu(\varepsilon)| = 1$ for all energies in the range $[V_- - m, V_+ + m]$ where the solution inside the barrier is non-oscillatory and $\alpha$ is real; otherwise, $|\mu(\varepsilon)| < 1$. For completeness, we provide the two-component wavefunction amplitudes given by

$$A_+(\varepsilon) = \tfrac{1}{2\alpha} [(i\gamma + \alpha) - (i\gamma - \alpha) R(\varepsilon)], \quad (14a)$$
$$B_+(\varepsilon) = \tfrac{-1}{2\alpha} [(i\gamma - \alpha) - (i\gamma + \alpha) R(\varepsilon)], \quad (14b),$$

while $A_- = \alpha A_+$ and $B_- = -\alpha B_+$.

For $S < -m$ the positive/negative energy solutions inside the barrier are given by (7a)'/(7b)'. Moreover, sub-barrier full transmission is possible only if the positive energy solutions outside the barrier overlap with the negative energy solution inside the barrier. That is, if we take $V_+ + m \geq m$; equivalently, $-m > S \geq -V$. Solving for the reflection and transmission amplitudes, we obtain the exact same results given by Eqs. (12) and (13).

As a result of studying the above two cases, $S > -m$ and $S < -m$, we conclude that sub-barrier full transmission is possible only if the strength of the scalar barrier is bound by $V - 2m \geq S \geq -V$. This, of course dictates that $V \geq m$. Moreover, transmission resonances occur at energies where $|R(\varepsilon)| = 0$ or, equivalently, $|T(\varepsilon)| = 1$. From (12) and our remark about $|\mu(\varepsilon)|$, we conclude that these resonances occur when the condition $e^{2pa} = 1$ or $\mu = 0$. The first condition takes place when $2pa = \pm 2in\pi$, where $n = 1, 2, 3,..$ (note that $n = 0$ is excluded). This means that resonant transmission occurs at the discrete energies $\varepsilon_n = V \pm \sqrt{(S+m)^2 + (n\pi/a)^2}$. The second condition implies a single resonance at $\varepsilon_0 = -\tfrac{V}{S} m$, which holds only for the case $0 > S \geq -V$ since $\varepsilon \geq m$ and $V \geq m$. Therefore, full transmission at zero momentum means that $\varepsilon_n = m$. This implies that zero momentum resonant transmission occurs for the following values of $S$:

$n = 0$: $\quad S = -V$, $\quad (15a)$

$n = 1, 2,..., \tfrac{a}{\pi}(V-m)$: $\quad S = -m \pm \sqrt{(V-m)^2 - (n\pi/a)^2}$. $\quad (15b)$

The first case, where $n = 0$, corresponds to the pseudo-spin symmetric case. The second case has a simultaneous resonant transmission at $\varepsilon = 2V - m$.

## IV. Results, discussion and conclusion

Figure 1 is a set of plots of the reflection and transmission coefficients for a given set of physical parameters corresponding to the cases given by (15a) and (15b). The figure clearly shows full transmission at $\varepsilon = m$ in addition to the usual above barrier resonances. On the other hand, Fig. 2 shows the same but for unconstrained potential strengths with $-m \geq S \geq -V$. The figure shows the interesting phenomenon of full transmission within a relatively very wide sub-barrier energy band in the Klein energy zone[†]. Figure 3 gives $|\mu(\varepsilon)|^2$ for the same problem associated with Fig. 2. Table I gives the resonance energies associated with Fig. 1 and Fig. 2. Figure 4 is a snap shot of a video animation for fixed values of $a$ and $V$ but with varying values of $S$ [17].

In conclusion we can say that super-critical transmission occurs for scalar and vector couplings including the pseudo-spin symmetric case where $S = -V$. This phenomenon is not present in other well-known potential configurations such as spin symmetric models where $S = V$. Our results also show that sub-barrier full transmission is possible only if the strength of the scalar barrier is bound by $V - 2m \geq S \geq -V$. Moreover, for coupling such that $-m \geq S \geq -V$ full transmission can occur within a wide sub-barrier energy band in the Klein energy zone. This result is completely a relativistic effect. In fact, it is worth mentioning in this context that between the two mode of coupling where $S = \pm V$, only the spin symmetric case, where $S = V$, produces a nontrivial non-relativistic limit [16]. The existence of such full transmission bands should not result in the failure of electron confinement in usual microelectronic devices since usual barrier heights in solid state devices are of the order of eV while critical barrier heights allowing for Klein tunneling are of the order of MeV. However, graphene systems whose current carriers are almost massless Dirac fermions [7] suggest that Klein tunneling is easily observed in such systems and hence renders future graphene technological applications a real challenge. The role played by the sharp boundaries of the square barrier will be investigated by considering a Woods-Saxon version of our potential and will be communicated in the near future.

**Acknowledgments:** The generous support provided by the Saudi Center for Theoretical Physics (SCTP) is highly appreciated by all Authors. We also acknowledge partial support by King Fahd University of Petroleum & Minerals under research group project RG1108-1 & RG1108-2.

---

[†] The Klein energy zone lies within the energy range bounded by m and the minimum of $(V_{\pm} \pm m)$.

**Table Caption**

**Table I**: Transmission resonance energies associated with the physical parameters of Fig. 1 and Fig. 2

Table I

| Fig. 1a | Fig. 1b | Fig. 1c | Fig. 2a | Fig. 2b |
|---|---|---|---|---|
| 1.0000000 | 1.0000000 | 1.0000000 | 1.6000000 | 1.7030917 |
| 5.5431086 | 5.0000000 | 1.3404935 | 1.8280421 | 2.5000000 |
| 6.7241918 | 6.3767149 | 5.0000000 | 6.1719579 | 3.1379041 |
| 8.1192392 | 7.8722899 | 6.3767149 | 7.4813222 | 6.8620959 |
| 9.5938166 | 9.4039844 | 7.8722899 | 8.9453625 | 8.2969083 |
|  |  | 9.4039844 |  | 9.8173239 |

**Figures Captions**

**Fig. 1**: The solid/dashed curve represents the transmission/reflection coefficient associated with the physical parameters $V = 3$, $a = 2$, $m = 1$, and the following choices of $S$:
 (a) $S = -V$ , (b) $S = -m + \sqrt{(V-m)^2 - (\pi/a)^2}$ , (c) $S = -m - \sqrt{(V-m)^2 - (\pi/a)^2}$
Resonant transmissions are indicated by the solid black squares on the energy axis whose values are shown in Table I. These include resonant tunneling at zero momentum.

**Fig. 2**: The solid/dashed black curve is the transmission/reflection coefficient associated with the physical parameters $a = 2$, $m = 1$, and the following choices of $V$ and $S$:
 (a) $V = 4.0$ and $S = -2.5$, (b) $V = 5.0$ and $S = -2.0$
Full transmission is evident within a relatively very wide sub-barrier energy band in the Klein energy zone. Resonant transmissions are indicated by the solid black squares on the energy axis whose values are shown in Table I.

**Fig. 3**: Plots of the energy function $|\mu(\varepsilon)|^2$ corresponding to the same physical parameters of Fig. 2a and Fig. 2b. It is evident that $|\mu(\varepsilon)| = 1$ for all energies in the range $[V_- - m, V_+ + m]$. Moreover, $|\mu(\varepsilon)| = 0$ at $\varepsilon = -\frac{V}{S}m$ (solid black square).

**Fig. 4** : Snap shot of a video animation of $|T(\varepsilon)|^2$ (blue) and $|R(\varepsilon)|^2$ (red) for fixed values of $a$ and $V$ but with varying values of $S$. We took $V = 5$, $a = 2$, $m = 1$, and $S$ varies from $-6$ to $4$. The instantaneous values of $S$ is displayed in the video. The shot was taken at $S = -2.5$

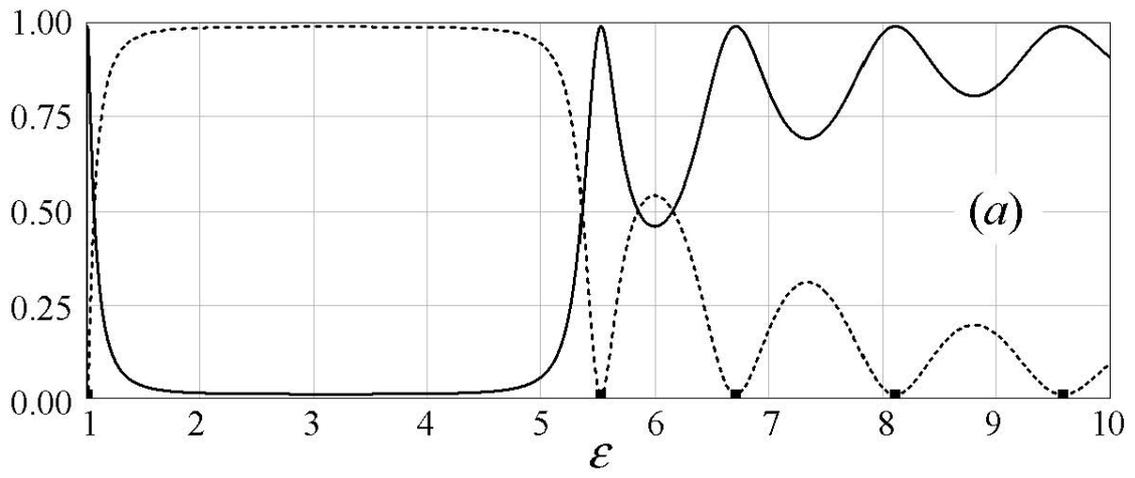

**Fig. 1a**

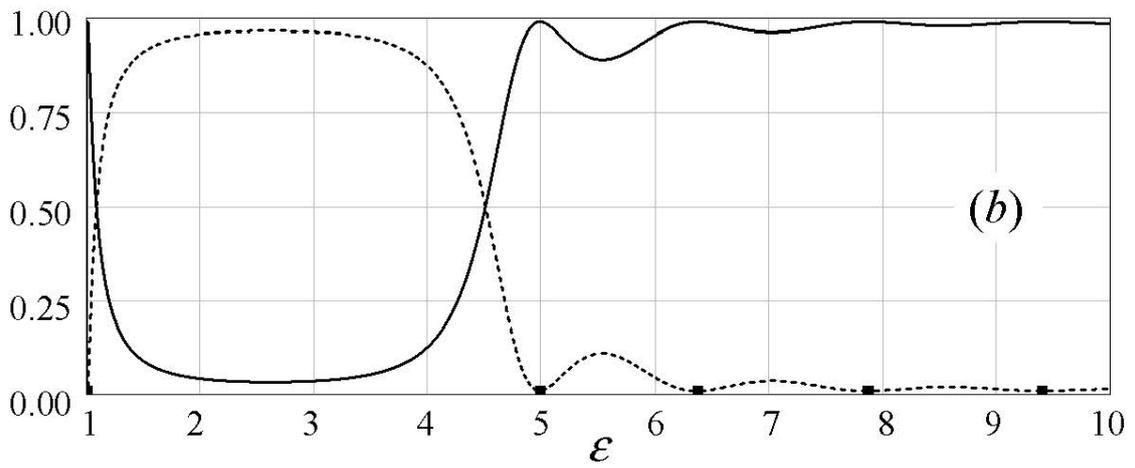

**Fig. 1b**

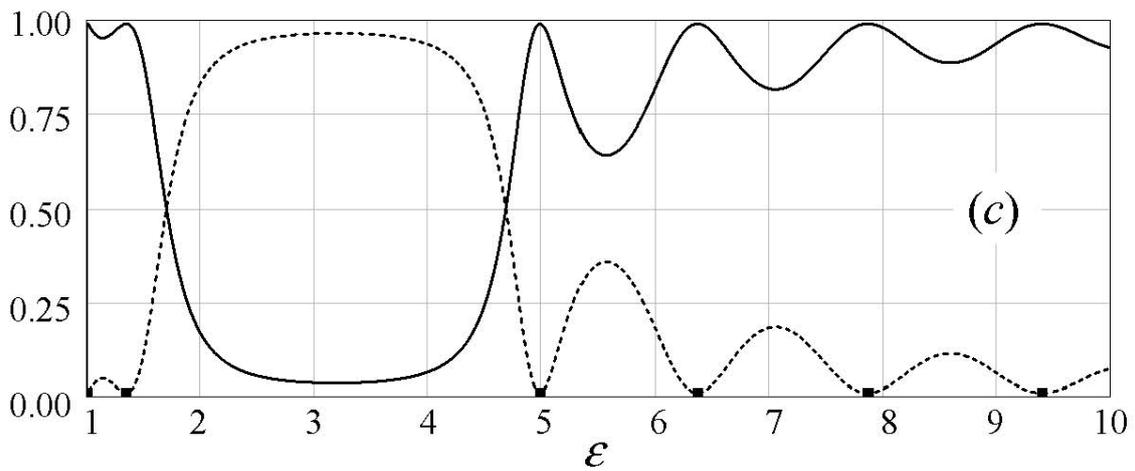

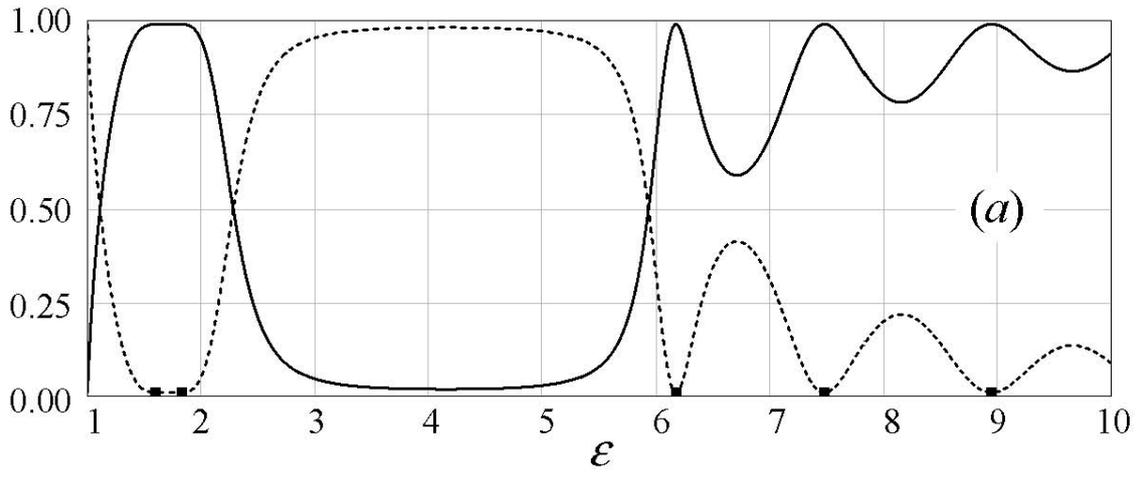

Fig. 2a

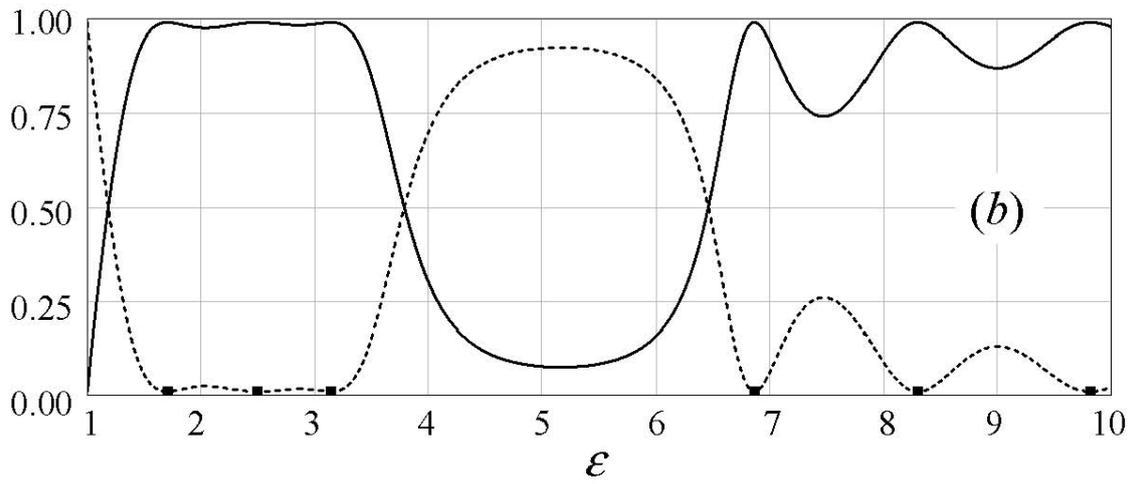

Fig. 2b

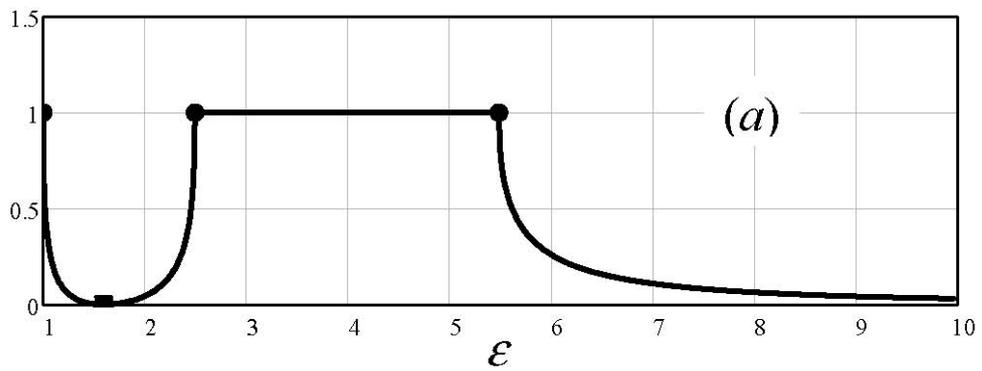

Fig. 3a

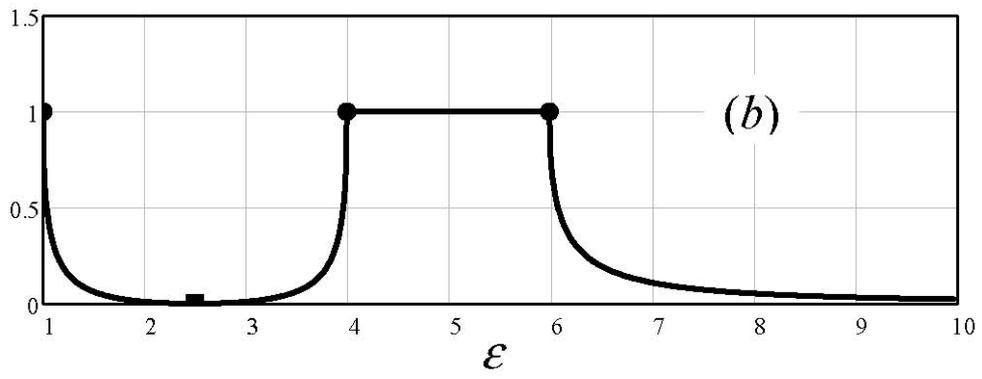

**Fig. 3b**

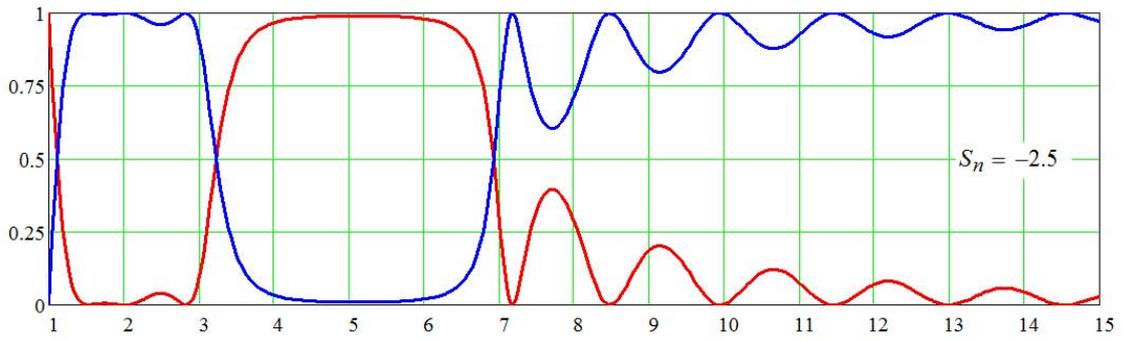

**Fig. 4**